\documentclass[12pt]{iopart}

\usepackage{epsfig}
\usepackage{graphicx} 
\begin{document}

\title[Nwankwo et al. ]{Assessment of Radio-Frequency Radiation Exposure Level from Selected Mobile Base Stations (MBS) in Lokoja, Kogi State, Nigeria}

% Use optional labels to link authors explicitly to addresses:
\author{Victor U.J. Nwankwo$^{1,3}$, Norbert N. Jibiri$^{2}$, Silas S. Dada$^{1}$, Abraham A. Onugba$^{1}$ and Patrick Ushie$^{1}$}
 \address{$^{1}$College of Natural and Applied Sciences, Salam University, Lokoja, Nigeria}
 \address{$^{2}$Radiation and Health Research Laboratory, Dept. of Physics, University of Ibadan, Nigeria}
 \address{$^{3}$S. N. Bose National Centre for Basic Sciences, Kolkata 700098, India}

%\author{}

%\address{IOP 
%Publishing, Dirac
%House, Temple Back, Bristol BS1 6BE, UK}
\ead{victornwankwo@yahoo.com}

\begin{abstract}
The acquisition and use of mobile phone is tremendously increasing especially in developing countries, but not without a concern. The greater concern among the public is principally over the proximity of mobile base stations (MBS) to residential areas rather than the use of handsets. In this paper, we present an assessment of Radio-Frequency (RF) radiation exposure level measurements and analysis of radiation power density (in $W/m^{2}$) from mobile base stations relative to radial distance (in metre). The minimum average power density from individual base station in the town was about $47\mu W/m^{2}$ while the average maximum was about $1.5mW/m^{2}$. Our result showed that average power density of a base station decreases with increase in distance (away from base station) and that radiation intensity varies from one base station to another even at the same distance away. Our result (obtained signature of power density variation from data) was also compared the with `expected'  signature. It was found that radiation from external sources (indicative) interfered with the reference base station and accounted for the deviation observed in this study. Finally, our result showed that the RF exposure hazard index in the town of Lokoja was below the permitted RF exposure limit to the general public recommended by ICRNIP. Useful recommendations were also made to the Policy and Regulatory Agencies responsible to Telephony in Nigeria.

\end{abstract}

\maketitle

\section{Introduction}
The acquisition and use of mobile phone for the purpose of communication has witnessed a tremendous increase across the world over the past two decades. The breakthrough in wireless telecommunication which enables mobile phone users to access the Internet (among other uses) via the mobile phone has even added weight to this unprecedented increase in GSM use. This has therefore led to a resultant increase in the installation of Mobile Base Stations (MBS) as service providers clamours for efficiency and quality service in order to compete favourably with their counterparts. There are over 1.6 billion mobile phone users worldwide. Mobile base station sites has also been estimated to be about 5 million worldwide, a number expected to grow to more than 11 million by 2020 [3]. Indeed, the Global System of Mobile (GSM) communication has proved to be of tremendous benefit to the society especially in a developing country like Nigeria, where other forms of communication exist to a very limited extent. However, associated with this seeming 'welcomed development' and 'life changing' technology is a concern about the health implication of the Electromagnetic radiation from both the mobile phone itself and their control-base stations that could result due to human exposure. The greater concern among the public is mainly over the proximity of base stations to both residential and official residence rather than the use of handsets.\\

Over the past one decade in Nigeria, mobile wireless services grew from niche market applications to nationally (and globally) available components of daily life. Lokoja town and its environs have over the years also experienced this upsurge in the use of GSM. The number of Mobile base stations in the town has also witnessed an exponential increase as a result. The city is therefore not left out of this growing concern, especially as buildings for residential purposes are seen being erected at close proximity to mobile base stations. Service providers are also mounting their base stations (and antennas) within and around public place such as schools, markets, and official residence without minding the health implication of this technology. More worrisome is the fact that there seem to be no precedence on which the use (and installation) of this technology is based. This informs the reason why research in this direction should be encouraged among scientists in Nigeria and other developing countries. In this study (preliminary) we estimate the exposure level of the public to RF radiations from mobile phone base station, particularly in Lokoja town and its environ.\\

One of the health effects from RF fields identified in scientific reviews has been related to an increase in body temperature ($< 1 ^{\circ} C$) from exposure at very high field intensity found only in certain industrial facilities, such as RF heaters [3]. There is also a concern about the effect of cumulative RF radiation resulting from continuous exposure. This have led to serious debates that this long-term EM radiation exposure may lead to some diseases like cancer and leukemia. Reported effects resulting from EM radiation also include symptoms of sleep-disorders, headaches, nervousness, fatigue, and concentration difficulties [1]. Also, many other studies have been made, both in epidemiological and experimental studies. The
Swiss Agency on Environmental, Forests and Landscape (SAEFL), published a report based
on more than 200 scientific studies carried out on human beings for the purpose of assessing
the health risks associated with the exposure to high frequency non-ionizing radiations [4]. In
Germany T-Mobile commissioned ECOLOG, and reviewed over 220 pieces of peer reviewed
and published papers in which they found evidences for effects on the central nervous
system, cancer initiating and promoting effects, impairment of certain brain function, loss of
memory and cognitive function [5].\\

Many scientific organizations, such as Institute of Electrical and Electronics Engineers (IEEE), International Commission on Non-Ionizing Radiation Protection (ICNIRP), National Radiation Protection Board (NRPB), and the Federal Communication Commission (FCC) of the United States of America, have made efforts to protect occupationally exposed persons and general public against the adverse effects of radio-frequency (RF) radiations, by setting up different guidelines. The ICNIRP guidelines have been widely adopted by many countries and the limits set are used as national RF safety standards. The permitted level for RF exposure to the general public at a frequency of 900 MHz is 4.5 $W/m^{2}$ (power density) or 41 V/m (electric field strength) and at 1800 MHz we have 9 $W/m^{2}$ (power density) or 58 41 V/m (electric field strength). The same levels are also recommended by the United Nation's World Health Organisation (WHO) [6].

\subsection{Radio Frequency (RF) emission in the Environment}
Radio frequency (RF) emission is one of several types of electromagnetic fields (EMF). Electromagnetic energy consists of waves of electric and magnetic energy components moving together through space at the speed of light. The EMF is a term used to describe energy that travels through air or space, the most common form of which is visible light. We come into daily contact with EMF in many different forms. Other sources of EMF include televisions, microwave ovens, computers, light bulbs, cordless phones and digital clock radios. The movement of electrical charges generates these waves. In the case of cellphone technology, the movement of charges (i.e alternating current)in a transmitting radio antenna creates electromagnetic waves that is emitted away from the antenna and can be picked up by a receiving antenna. The electromagnetic spectrum is shown in figure 1. The electromagnetic spectrum covers an enormous range of frequencies. The electromagnetic waves specific to cellphone technology, are known as non-ionising radiation. This implies that they are not capable of breaking chemical bonds in biological structures (such as humans) or removing electrons (ionisation). Figure 2 shows the typical power of the radio services in the community when transmitting.

\begin{figure}
\label{figure1}
\begin{center}
\includegraphics*[width=10cm,]{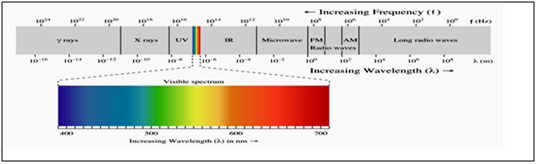}
\end{center}
\caption{The electromagnetic spectrum}
\end{figure}

\begin{figure}
\label{figure2}
\begin{center}
\includegraphics*[width=10cm,]{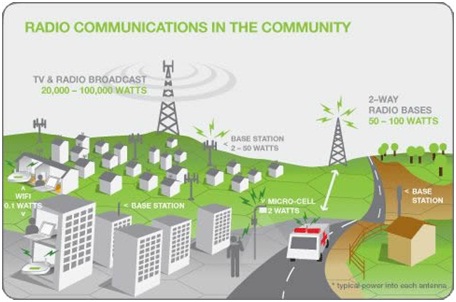}
\end{center}
\caption{Radio communications in the community\\
(Source http://www.emfexplained.info/?ID=25186)}
\end{figure}

\subsection{Study Area}
The study area for this project is Lokoja. Lokoja (town and river port) is the capital of Kogi State, north-central Nigeria, on the west bank of the Niger River opposite the mouth of the Benue River [7]. This place is situated in Kotonkar, Kogi, Nigeria, its geographical coordinates (location) are 7$^\circ$ 48' 0" North, 6$^\circ$ 44' 0" East at an elevation/altitude of meters. The average elevation of Lokoja, Nigeria is 55 meters and its original name (with diacritics) is Lokoja [8]. According to the 2006 population census, Lokoja has a population of 195,261 (National Bureau of Statistics, Nigeria). It covers an area of about 90 $km^{2}$ and with a population density of 70/$Km^{2}$. It is bounded by the Niger in the north and east upstream from the capital until the border with Kwara State. The satellite image/map of Lokoja and its neighbouring towns is shown in the figure 3.

\begin{figure}
\label{figure3}
\begin{center}
\includegraphics*[width=10cm,]{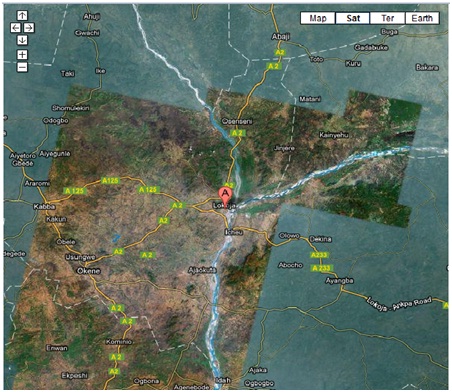}
\end{center}
\caption{Satellite image of Lokoja and its environ}
\end{figure}

\section{Materials and Method}
In this work the method of broadband analysis was employed in the measurements. A hand held TES-90 Electrosmog meter was used in the RF survey. The meter is a Broad band device for monitoring high frequency radiation in the range of 50 MHz to 3.5 GHz. The meter measures the value the electric field {\bf E} and converts it into the magnetic field {\bf H} and the power density {\bf S} using equation (1)[6]. This is the power density {\bf S} (e.i the power per unit area) expressed in Watts per Meter squared ($W/m^{2}$). The meter can measure {\bf E} along different axis, but can also take readings in all {\bf Es} at the same time (Triaxial) using equation (2).

\begin{eqnarray}
S = EH = \frac{E^2}{377} = 377\Omega H^{2}
  \end{eqnarray}
  
\begin{eqnarray}
E^{2} = \mathrm{E}_x^2 + \mathrm{E}_y^2 + \mathrm{E}_z^2
  \end{eqnarray}

Measurements (of radiation power density) were made by simply pointing the meter to the source of the RF radiation. A maximum of about 150m radial distance from the foot of the base station was considered and measurement were taken  at 25m interval from each base station. The proximity of residential buildings to (and manner in which structures were erected around) base stations informed the reason for choosing this range of distance. A total of about 16 base stations were randomly selected in Lokoja town,  (covering about 5 network providers - Globacom, MTN, Etisalat, Multilink and Airtel), according to their proximity to buildings, number of antennas mounted on their masts, how close they are to other base stations and the population density around them. The meter was set to the triaxial measurement mode and also to the maximum instantaneous measurement mode, to measure the maximum instantaneous power density at each point. Each measurement was made by holding the meter away from the body, at arm's length and at about 1.5m above the ground level pointing towards the mast as suggested by Ismail et al (2010). The values of the measured power densities taken after the meter is stable (about 3 min) were recorded. We ensured that the measured values were not influenced by unwanted sources and disturbances. Such precautions taken were to avoid the movement of the meter during measurements and excessive field strength values due to electrostatic charges, and to make sure (where possible) that movement of cars and phone calls were reduced before taking measurements. Using a software package, the geographical location of surveyed (selected) base stations (listed in table 1) were located on the map of Lokoja and its environs and shown in the figure 4.

\begin{figure}
\label{figure4}
\begin{center}
\includegraphics*[width=10cm,]{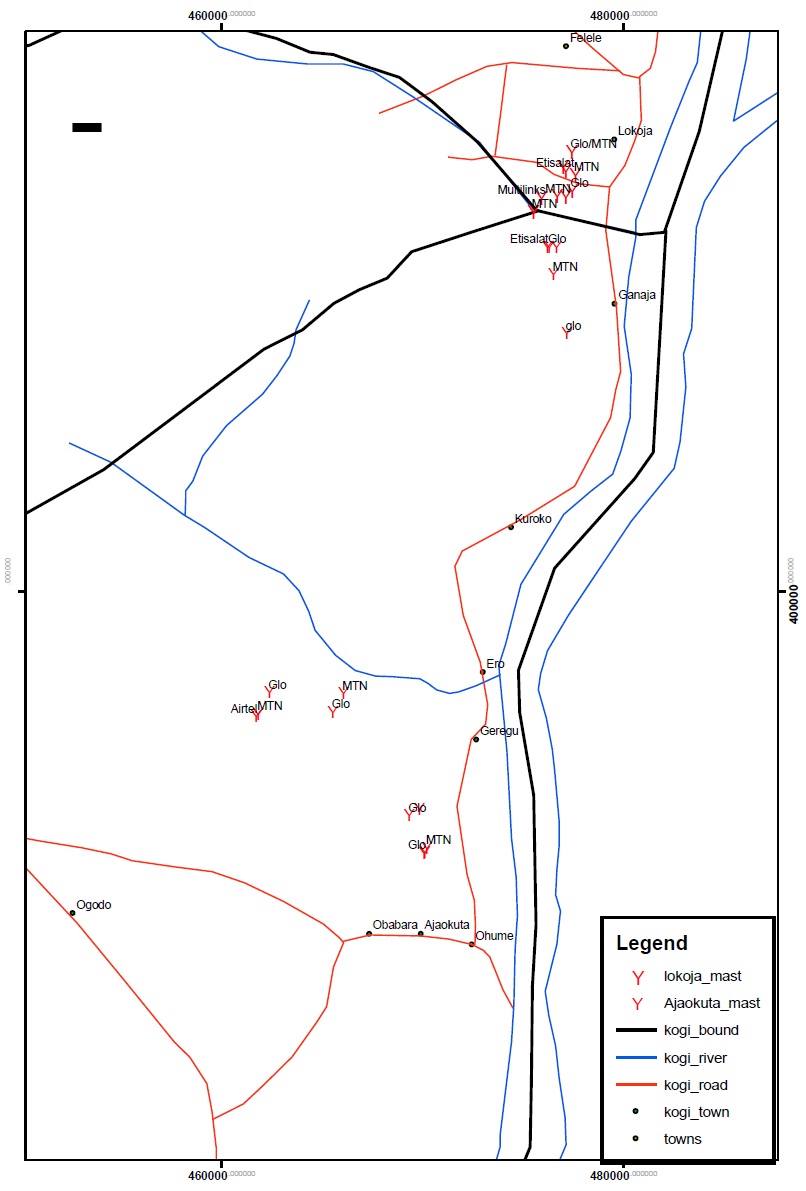}
\end{center}
\caption{Selected base stations located on the map}
\end{figure} 

\begin{table}
\caption{\label{arttype} Position of surveyed (selected) mobile stations in Lokoja}
\footnotesize\rm
\begin{tabular*}{\textwidth}{@{}l*{15}{@{\extracolsep{0pt plus12pt}}l}}
\br
Base Station& &Position/location\\
\mr
\verb"MBS1"& &Lat 07$^\circ 44'N$, Long 006 $^\circ 44.7'E$\\
\verb"MBS2"& &Lat 07$^\circ 46'N$, Long 006$^\circ 43.4'E$\\
\verb"MBS3"& &Lat 07$^\circ 46.4 'N$, Long 006$^\circ 44.4'E$\\
\verb"MBS4"& &Lat 07$^\circ 46.3'N$, Long 006$^\circ 44.3'E$\\
\verb"MBS5"& &Lat 07$^\circ 46.35'N$ Long 006$^\circ 46.2'E$\\
\verb"MBS6"& &Lat 07$^\circ 47.7'N$ Long 006$^\circ 44'E$\\
\verb"MBS7"& &Lat 07$^\circ 47.75'N$ Long 006$^\circ 44.45'E$\\
\verb"MBS8"& &Lat 07$^\circ 47.72'N$ Long 006$^\circ 44.69'E$\\
\verb"MBS9"& &Lat 07$^\circ 47.9'N$ Long 006$^\circ 44.83'E$\\
\verb"MBS10"& &Lat 07$^\circ 48.34'N$ Long 006$^\circ 44.95'E$\\
\verb"MBS11"& &Lat 07$^\circ 48.45'N$ Long 006$^\circ 44.7'E$\\
\verb"MBS12"& &Lat 07$^\circ 48.54'N$ Long 006$^\circ 44.7'E$\\
\verb"MBS13"& &Lat 07$^\circ 48.97'N$ Long 006$^\circ 44.8'E$\\
\verb"MBS14"& &Lat 07$^\circ 47.3'N$ Long 006$^\circ 43.8'E$\\
\verb"MBS15"& &Lat 07$^\circ 33.48'N$ Long 006$^\circ 36.38'E$\\
\verb"MBS16"& &Lat 07$^\circ 33.61'N$ Long 006$^\circ 38.45'E$\\
\br
\end{tabular*}
\end{table}

\section{Results and Discussion} 
We observed that the average power density of MBS6 was practically the highest compared to that of the remainder, with a contribution of about $20\%$ ($1.4mW/m^{2}$) as observed in table 3, fig 5 and 6. The least contribution comes from MBS3 with only about $1\%$ ($47\mu W/m^{2}$). Others with significant contribution are MBS1 ($12\%$), MBS5 ($10\%$) and MBS10 ($10\%$). Their respective power densities are 865.9 $\mu W/m^{2}$, 740.48 $\mu W/m^{2}$ and 737.27 $\mu W/m^{2}$. We also observed a significant fluctuation in data during measurement. One would have expected a decrease in power density by the square of radial distance ($P_t/4\pi R^{2}$)  as you move farther away from reference base station; this was however not so in most cases as observed in table 2. Fluctuations can be attributed to one or more of the five factors observed during measurement: (i) obstruction constituted by immobile structures placed or erected within the line of sight of measurement (ii) wave interference from other sources of electromagnetic radiation around reference base station such as radio and TV antennas, receivers etc. (iii) interference from radiation and/or noise from moving objects such as vehicles, motorcycles etc. (iv) topography (or elevation) of the land area around reference base station with respect to radial distance away from base station and (v) wave interference from other mobile base stations clustered around a reference base station. In general, the mean power density (combined) decreased steadily with increase in radial distance as observed in figure 5 (a), (b) and (d). Figure 5c is the ‘expected’ signature or plot (but not from the analyzed data) of variation in power density with distance. 

\begin{table}
\caption{\label{arttype} Measured power density of the MBS at 25m interval}
\footnotesize\rm
\begin{tabular*}{\textwidth}{@{}l*{15}{@{\extracolsep{0pt plus12pt}}l}}
\br
Base Station& &Distance(m)\\
 &25&50&75&100&125&150\\
\mr
\verb"MBS1"&408.00&305.00&135.00&115.80&194.00&100.80\\
\verb"MBS2"&944.10&2488.00&810.80&549.30&245.10&158.10\\
\verb"MBS3"&30.70&38.10&2.30&11.20&107.10&93.40\\
\verb"MBS4"&839.80&487.20&365.60&324.50&164.50&103.30\\
\verb"MBS5"&979.80&888.00&694.20&692.80&681.00&507.10\\
\verb"MBS6"&1267.00&1900.00&1927.00&1700.00&1120.00&1020.00\\
\verb"MBS7"&451.70&281.80&426.30&929.60&659.60&551.00\\
\verb"MBS8"&591.40&318.60&142.60&290.70&205.80&155.50\\
\verb"MBS9"&705.40&186.20&128.00&103.20&90.10&70.60\\
\verb"MBS10"&1506.00&874.20&956.00&784.90&173.20&129.30\\
\verb"MBS11"&828.20&1020.00&602.70&227.20&288.50&208.50\\
\verb"MBS12"&264.70&217.70&192.70&70.00&60.40&40.80\\
\verb"MBS13"&1549.00&817.60&303.70&203.40&200.70&178.50\\
\verb"MBS14"&561.20&366.80&595.50&488.10&450.00&433.10\\
\verb"MBS15"&289.90&164.40&150.50&128.90&100.80&72.10\\
\verb"MBS16"&478.00&390.10&249.50&120.10&84.60&7.90\\
\br
\end{tabular*}
\end{table}

\begin{table}
\caption{\label{arttype} Mean power density of each MBS and their percentage contribution ($\%$) contribution}
\footnotesize\rm
\begin{tabular*}{\textwidth}{@{}l*{15}{@{\extracolsep{0pt plus12pt}}l}}
\br
Base Station&Mean power density (in $\mu W/m^{2}$)  &$\%$ contribution\\
\mr
\verb"MBS1"&209.92&3.0\\
\verb"MBS2"&865.90&12.0\\
\verb"MBS3"&47.13&1.0\\
\verb"MBS4"&380.82&5.0\\
\verb"MBS5"&740.48&10.0\\
\verb"MBS6"&1489.00&20.0\\
\verb"MBS7"&550.00&7.0\\
\verb"MBS8"&284.10&4.0\\
\verb"MBS9"&213.92&3.0\\
\verb"MBS10"&737.27&10.0\\
\verb"MBS11"&529.18&7.0\\
\verb"MBS12"&141.05&2.0\\
\verb"MBS13"&542.15&6.0\\
\verb"MBS14"&482.45&6.0\\
\verb"MBS15"&151.10&2.0\\
\verb"MBS16"&221.70&3.0\\
\mr
\verb" "&7586.17&100.0\\
\br
\end{tabular*}
\end{table}

\begin{table}
\caption{\label{arttype} Mean power density (combined) of all the surveyed MBS}
\footnotesize\rm
\begin{tabular*}{\textwidth}{@{}l*{15}{@{\extracolsep{0pt plus12pt}}l}}
\br
Base Station&Distance (m)&Mean power density (in $\mu W/m^{2}$)\\
\mr
\verb"MBS1-16"&25&730.93\\
\verb" "&50&671.54\\
\verb" "&75&480.15\\
\verb" "&100&421.06\\
\verb" "&125&301.76\\
\verb" "&150&239.38\\
\br
\end{tabular*}
\end{table}

\begin{figure}
\label{figure5}
\begin{center}
\includegraphics*[width=12cm,]{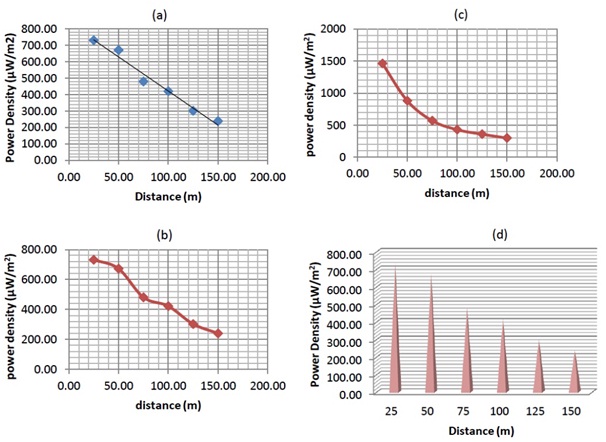}
\end{center}
\caption{Mean power density plot/signature of all MBS: (a), (b) and (d) are from data, (c) expected signature}
\end{figure} 

\begin{figure}
\label{figure6}
\begin{center}
\includegraphics*[width=10cm,]{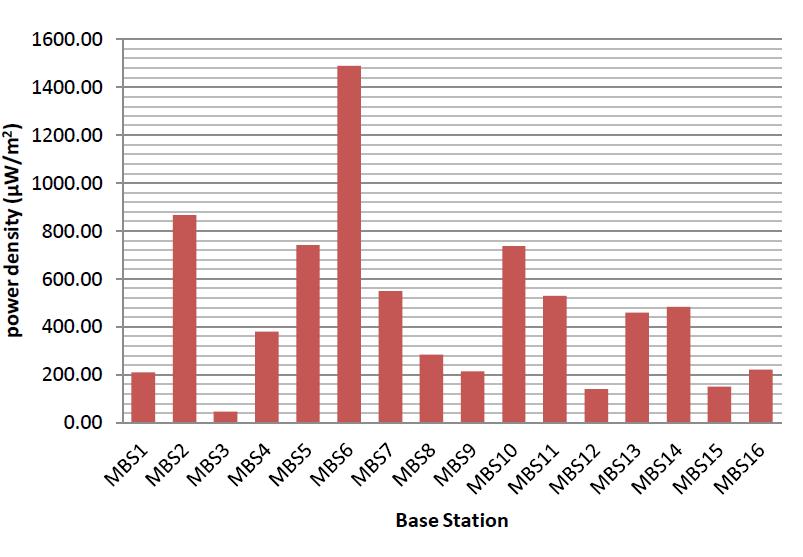}
\end{center}
\caption{Average power density of each base station}
\end{figure} 

\begin{figure}
\label{figure7}
\begin{center}
\includegraphics*[width=10cm,]{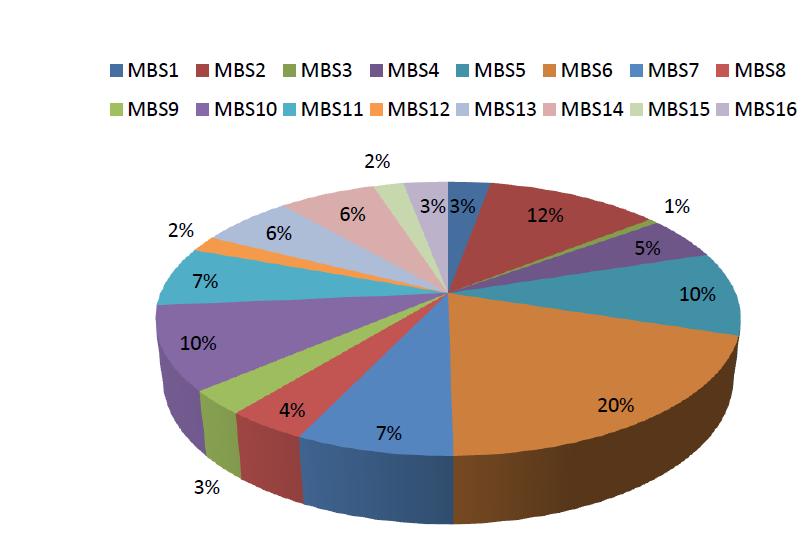}
\end{center}
\caption{Percentage contribution of each base station}
\end{figure}

\section{Conclusion} 
The mean (average) power density of base station decreased as the radial distances (away from the base stations) were increased, and radiation intensity varies from one mobile base station to another (even at the same distance away).We observed that Mobile base stations with little or no fluctuation (in power density with distance) have little or minimal interference from external sources, while those with noticeable data fluctuations have significant interference from external sources. Fluctuations were due to one or more of the five factors observed during measurement: (i) obstruction constituted by immobile structures placed or erected within the line of sight of measurement (ii) wave interference from other sources of electromagnetic fields around reference base station, such as radio and TV antennas, receivers etc. (iii) interference from radiation (wave) and/or noise from moving objects such as vehicles, motorcycles etc. (iv) topography (or elevation) of the land area around reference base station with respect to radial distance away from base station and (v) wave interference from other mobile base stations clustered around a reference base station. The minimum average power density from individual base station in the town was about 47$\mu W/m^{2}$ while the mean maximum was about 1.5$mW/m^{2}$. Therefore the RF exposure hazard index in the town of Lokoja was below the permitted RF exposure limit to the general public recommended by ICRNIP. However, as much as possible, mobile network providers should site mobile base stations at least 150m away from residential areas and other sources of electromagnetic radiation.	Mobile base stations whose RF radiation intensity is considerably high should be identified and settlers should be made to relocate farther away from such base stations. Government should inert and enforce a law that ensures that service providers adheres strictly to set precedence/guidelines for installation of mobile base station, especially relating to their proximity to residential and official areas. In view of the potential hazards of long-term exposure, national regulatory agencies should in collaboration with other government agencies (at both state and federal level) set up and embark on enlightenment campaign at regular interval to sensitize the public on the implications of RF radiation exposure and safety measures. Government at both state and federal level should enforce and insist on town planning as a matter of utmost importance in Lokoja town and its environs, while providing an enabling environment for builders to erect structures accordingly.

\section*{Acknowledgment}
We are thankful to Salem University, Lokoja, Nigeria for funding this research, and to Ayinmode Bolaji of the Department of Physics, University of Ibadan, Nigeria for his useful contribuions.  We also thank Afolabi Alex, Rueben O. Ajodo and Ngozi G. Ibekwete (all of CNAS, Salem University, Lokoja) for their efforts during this work.

\end{document}